%% file: paper.tex
\documentclass{article}

\usepackage{PRIMEarxiv}

\usepackage[utf8]{inputenc} 
\usepackage[T1]{fontenc}    
\usepackage{hyperref}       
\usepackage{url}            
\usepackage{booktabs}       
\usepackage{amsfonts}       
\usepackage{nicefrac}       
\usepackage{microtype}      
\usepackage{lipsum}
\usepackage{fancyhdr}       
\usepackage{graphicx}       
\graphicspath{{media/}}     

\usepackage[utf8]{inputenc}
\usepackage{graphicx}

\usepackage[english]{babel}
\usepackage{blindtext}
\usepackage[T1]{fontenc}
\usepackage{setspace}

\usepackage{subcaption}  
\usepackage{graphicx}
\usepackage{multirow}
\usepackage[inline]{enumitem}
\usepackage{float}
\usepackage{amssymb}  
\usepackage{amsmath}
\usepackage{hyperref}
\usepackage{cleveref}
\usepackage{makecell}
\usepackage{caption}
\usepackage[list=true,listformat=simple]{subcaption}
\usepackage[]{xcolor}

\newcommand{\ignore}[1]{}
\newcommand{\etal}{{et al.}}
\usepackage[normalem]{ulem}

\title{IRR-Based AS Type of Relationship Inference}

\author{
  Amit Zulan \\
  School of Electrical and Computer Engineering \\
  Tel-Aviv University \\
  \texttt{zulan@eng.tau.ac.il} \\
   \And
   Omer Miron \\
  School of Electrical and Computer Engineering \\
  Tel-Aviv University \\
  \texttt{omermiron@mail.tau.ac.il} \\
   \And
   Tal Shapira\\
  School of Computer Science \\
  The Hebrew University of Jerusalem \\
  \texttt{talshapirala@gmail.com} \\
   \And
  Yuval Shavitt \\
  School of Electrical and Computer Engineering \\
  Tel-Aviv University \\
  \texttt{shavitt@eng.tau.ac.il} \\
}

\begin{document}
\maketitle

\begin{abstract}
The Internet comprises tens of thousands of autonomous systems (ASes) whose commercial relationships are not publicly announced. The classification of the Type of Relationship (ToR) between ASes has been extensively studied over the past two decades due to its relevance in network routing management and security.

This paper presents a new approach to ToR classification, leveraging publicly available BGP data from the Internet Routing Registry (IRR). We show how the IRR can be mined and the results refined to achieve a large and accurate ToR database.  Using a ground truth database with hundreds of entries we show that we indeed manage to obtain high accuracy. About two-thirds of our ToRs are new, namely, they were not obtained by previous works, which means that we enrich our ToR knowledge with links that are otherwise missed.

\end{abstract}



\maketitle

\input{body.tex}

\bibliographystyle{unsrt}  
\bibliography{bibliography.bib}  






\end{document}

%% file: body.tex


\section{Introduction}
The
Internet is a vast collection of independently operated networks, known as Autonomous Systems (ASes), which include entities such as Internet Service Providers (ISPs), enterprises, and universities. Each AS is assigned a unique Autonomous System Number (ASN) and announces one or more IP address prefixes (APs) via the Border Gateway Protocol (BGP). BGP update messages provide the full AS path needed to reach a given AP.


In her foundational work~\cite{Gao}, Gao categorized the commercial relationships between interconnected ASes into three distinct types: \begin{enumerate*} \item Provider-to-customer (P2C) – where the customer AS compensates the provider AS for transit traffic to and from the broader Internet, \item Peer-to-peer (P2P) – where two ASes exchange traffic with each other and their customers without forwarding traffic to or from their respective providers or other peers, and \item Siblings (S2S) – where two ASes operate under the same administrative control. \end{enumerate*}.
Gao also established a routing principle known as Valley-Free (VF), which dictates that once traffic descends from a provider to a customer, it cannot return to a provider, aligning with conventional BGP routing configurations.


ToR information helps predict the routing paths chosen by BGP, particularly in scenarios involving link failures \cite{DJMS06}. It also helps uncover malicious interference in the routing infrastructure, known as IP hijack attacks~\cite{ChinaHijack,bgp2vec_hijack,bgphij}.

Due to the limited public availability of ToR information, numerous studies have focused on methods for its inference.
Most of these solutions~\cite{Gao,SARK,AToR,CAIDA07,NDToR,ToR_AnNet} are based on publicly available BGP announcement databases~\cite{RV} with some additionally incorporating data from the Internet Routing Registry (IRR)~\cite{irr}.  

The IRR is a decentralized database ISPs use to disclose their routing policies and peering arrangements. As contributions to the database are voluntary, it remains partially populated, prone to errors~\cite{du2022irr}, and not regularly maintained.  

Xia and Gao \cite{pte04} were the first to use the BGP Community Attribute, AS-SET objects, and IRR routing policies to deduce AS relationships. They looked for instances that contained the strings of "customer," "provider," or "peer" in their AS-SET object, and also for cases where AS relationships are explicitly described in the routing policies.  Both approaches are quite restrictive, and as a result, they
only obtained partial AS relationships (14\% of total AS pairs on Oct.\ 2003).
Siganos and Faloutsos~\cite{siganos2004analyzing} also suggested to use the import/export fields in the IRR for identifying ToRs. 
Dimitropoulos \etal~\cite{CAIDA07} used the organization description records IRR, but only to infer S2S relationships.
 
All the three publications above  \cite{pte04,CAIDA07,siganos2004analyzing}  noted that the IRR database is problematic to use for ToR classification since the data there is not always accurate, and it is hard to estimate the accuracy of each record.  As this is accepted by most of the community, the rich ToR data in the IRR is not used, and as we show many ToRs are missed.

In this paper, we show that IRR records can be mined with high accuracy. We use multiple heuristics and examine the classification from both ends of the AS link.  This allows us to assess the reliability of our heuristic, which is defined by the agreement of both side of the ToR, per each IRR object (entities) and filter out objects where our heuristic fails.  The remaining objects demonstrate very high reliability, as demonstrated by a comparison between heuristics.  A ground truth datasets with hundreds of ToRs show that indeed our heuristics reach high accuracy of 93.7\%.

What is significant about our results is that we obtain new ToRs that are not present in any of the previous datasets. After filtering, we obtain 194,397 ToRs, 66.38\% of them, namely 129,041 are new.  If we accept lower reliability and use the unfiltered dataset: we obtain 231,151 new ToRs from 316,906 classified ToRs (72.94\%).





Similar to Dimitropoulos \etal~\cite{CAIDA07} we also extract sibling relationships, but we use more fields of the IRR objects, and as a result, we obtained significantly more relationships: 24,743 siblings.
We compared the sibling relationships results with two datasets: ProbeLink \cite{coretoleaf_19} and a proprietary  dataset from BGProtect, which contains a collection of siblings relationships manually tagged by expert analysts.  2,020 of these ToRs appear in the ProbLink dataset, but only 794 appear as siblings (39.3\%).  The overlap with the BGProtect dataset has 1,029 ToRs, 739 appear as siblings (71.8\%).  This may suggest that ProbLink misidentifies many siblings.

The remainder of this paper is structured as follows. Section~\ref{sec:related} provides an overview of related work. In Section~\ref{sec:methods}, we introduce our heuristics, followed by a discussion of the inferred ToRs in Section~\ref{sec:results}. Section~\ref{sec:eval} evaluates our approach using two manually labeled datasets, and
details our experimental results, comparing them with prior methods and publicly available datasets. Lastly, we conclude the paper in the final section. 


\section{Related Work}\label{sec:related}

Extensive research has focused on inferring the Type of Relationship (ToR) between Autonomous Systems (ASes). Previous studies have largely relied on heuristic algorithms that extract information from BGP announcements or reconstruct AS-level routes using traceroute data.

Gao~\cite{Gao} conducted the first in-depth study on AS relationship inference. She introduced heuristic algorithms that deduce AS ToRs from BGP routing announcements, leveraging the observation that a provider AS typically has a higher graph degree than its customers, while peers have similar degrees. The algorithm identifies the top provider in each path and classifies ToRs in accordance with the valley-free routing principle.

Subramanian \etal~\cite{SARK} formulated the ToR maximization problem, which involves assigning labels to all edges in an undirected AS graph to optimize the number of valley-free paths within a set of BGP routes. Their approach analyzes AS graph fragments from multiple vantage points, ranking each AS using a reverse-pruning technique. The ToR classification is then derived by comparing rank vectors, with closely ranked ASes categorized as P2P, while others are identified as C2P.

Xia and Gao~\cite{pte04} utilized the BGP Community Attribute, AS-SET objects, and routing policies from IRR Databases to infer AS relationships. However, their approach had limitations, identifying only 14\% of AS pairs as of October 2003. They also found that both GAO~\cite{Gao} and SARK~\cite{SARK} were ineffective in detecting P2P relationships, a conclusion that aligns with our observations for SARK.

Battista \etal~\cite{BPP} established that the ToR optimization problem~\cite{Gao,SARK} is NP-complete and reformulated it into the ToR-D problem, which allows for a limited number of invalid paths. They further reduced ToR-D to 2SAT and introduced a heuristic algorithm for determining ToRs.

Cohen and Raz~\cite{AToR} proposed the Acyclic Type of Relationship (AToR) problem, which seeks to minimize invalid paths while ensuring that the directed graph remains acyclic. They addressed this by developing a heuristic algorithm for solving the K-AToR problem.

Dimitropoulos \etal~\cite{CAIDA07} applied IRR~\cite{irr} data to infer S2S relationships and refined the problem by recognizing that AS paths do not always adhere to a strictly hierarchical structure when inferring P2C and P2P relationships. Their method incorporated a reachability metric, sorted ASes accordingly, and grouped those with equivalent values into hierarchical levels. Their approach achieved classification accuracies of 96.5\% for C2P, 82.8\% for P2P, and 90.3\% for S2S relationships.

Shavitt \etal~\cite{NDToR} developed a near-deterministic approach for ToR inference (ND-ToR) to reduce dependence on heuristic methods. Their method constructs the Internet’s core as a subgraph of high-tier ASes, generated using three distinct techniques: the Greedy Max Clique (GMC) core~\cite{jellyfish}, the $k$-Core based on $k$-shell decomposition~\cite{kshell}, and the CAIDA Peers Core (CP)\cite{caida_asrank}, which represents the largest connected component of a P2P graph consisting of major tier-1 ASes. To classify the remaining links, they employed a three-stage process incorporating the valley-free rule and the k-shell index\cite{kshell} of adjacent ASes. Their approach successfully inferred approximately 58,000 ToRs with an accuracy exceeding 95\%, leveraging AS-level path data from RouteViews~\cite{RV} and DIMES~\cite{DIMES}.

Ruan and Susan Varghese~\cite{ruan} introduced a method similar to ND-ToR~\cite{NDToR}. Their algorithm first processes AS paths that include a Tier-1 AS. In the next phase, it resolves undetermined AS paths by identifying the top provider AS as the one closest to a Tier-1 AS. Finally, a voting mechanism is applied to infer any remaining ambiguous relationships.

Luckie \etal~\cite{CAIDA13} developed the AS-Rank algorithm to infer C2P and P2P links using BGP data. Their approach is based on three key assumptions:
\begin{enumerate*}
\item a clique of major transit providers exists at the top of the AS hierarchy,
\item most customers establish transit agreements to ensure global reachability, and
\item the absence of C2P link cycles is necessary for routing convergence.
\end{enumerate*}
Using these principles, they designed an algorithm to infer the customer cone of an AS, which is the set of ASes it can reach via P2C links—and demonstrated strong performance in their evaluations.

To overcome challenges in complex inference scenarios—such as non-valley-free routing, limited visibility, and unconventional peering practices, Jin \etal~\cite{coretoleaf_19} identified key interconnection features and introduced ProbLink, a probabilistic algorithm that demonstrated a lower error rate than AS-Rank~\cite{caida_asrank} in their validation. To enhance inference accuracy, their approach integrates additional data sources, including sibling relationships, BGP communities, and IXP information.

Shapira and Shavitt~\cite{ToR_AnNet} proposed BGP2VEC, a method that treats BGP route announcements as sequences, embedding each Autonomous System Number (ASN) into a vector representation that captures its latent characteristics. The ToR between two ASes is inferred from the difference between their ASN vectors, and their analysis showed that using the nearest ToR vectors yields an accuracy exceeding 92\%. This technique can be integrated with other approaches to infer ToRs in cases where data is incomplete, or confidence levels are low.

AS relationship may vary across peering points and APs \cite{CAIDA07}, what is termed as complex ToRs.  Neudorfer \etal~\cite{Neudorfer2013} were the first to suggest a method to infer these.  They suggested examining ToRs at the PoP to PoP level but managed to find only a few complex ToRs.  Later, Giotsas \etal~\cite{giotsas2014inferring} suggested a more elaborate algorithm that is based on BGP, traceroute, and geolocation data to infer a few thousand complex ToRs.  In this work, we disregard complex ToRs since their number is negligible.

\section{Method}\label{sec:methods}

In this section, we describe heuristics to extract C2P, P2P, and sibling ToRs from IRR.   

\subsection{IRR Structure}\label{sec:irr_structure}

The IRR is written in a specification language called Routing Policy Specification Language (RPSL). The current list (from 2021) of routing registries contains 23 registries as listed in \cite{irr}, we use them all.

\begin{figure}[h]
    \centering
    \includegraphics[width=0.825\linewidth]{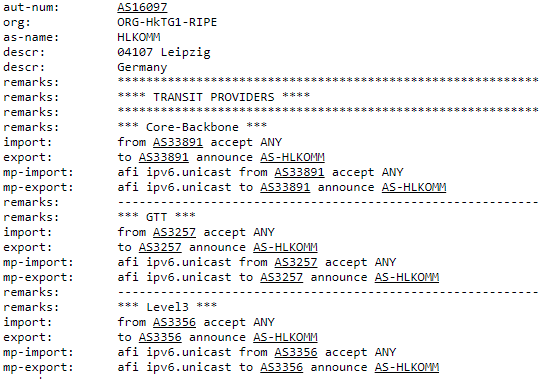}
    \caption{An AUT-NUM object example.}
    \label{fig:irr_block_example}
\end{figure}

RPSL is based on database objects (termed classes in the RFC 2622~\cite{RFC2622}); each contains some routing policy information and some necessary administrative data. We mainly use the AUT-NUM object (see Figure~\ref{fig:irr_block_example}), but also the AS-SET object.

The AUT-NUM object gives information about an AS. It contains mandatory fields such as aut-num (the ASN), as-name (The AS name) and optional fields containing peering information that can be used to deduct ToRs.
The {\em import} and the {\em export} optional fields express the BGP import and export policies of the ASN, and the {\em remarks} field is used to divide the RPSL text into easy to identify sections in free form, e.g., in Figure~\ref{fig:irr_block_example} there is a remark line that specifies that the following import/export fields are for transit providers. 
An important optional field that is almost always available is {\em last-modified}, which helps in estimating the correctness of the information.
An AS-SET object defines a collection of AS Numbers that can be utilized in contexts where a single AS Number would normally be referenced.

\subsection{Data Extraction and Pre-processing}

We extracted the relevant information from AUT-NUM objects and AS-SET objects in all the registries listed in IRR~\cite{irr}.
In multiple blocks for the same AS or AS-SET, only the most up-to-date block across all registries is processed.

For each AS-SET object, we extract the set's name, members of the set, and the date in which this block was last updated.
AS-SETs often have members that are also AS-SETs. Therefore we recursively break down each set, replacing every non-numeric member with either the ASN (AS number) that matched the member name or the members of the set specified in the member's name. As mentioned above, the database is not supervised, and there are many cases of cyclic reference in AS-SETs. To deal with this issue, we remember the already encountered sets and ignore a second reference of the same set.

In the AUT-NUM analysis, the information extracted is the ASN, the AS name, the import and export policies, and the object's last update date. Then, wherever an AS-SET is encountered in any field, we replace it with its respective list of ASNs. Furthermore, we apply an analysis based on the 'remarks' fields which are described in Sec.~\ref{sec:irr_structure}.

\subsection{Heuristics}
\subsubsection{Import-Export (I/E) Heuristic}

Each AS announces its import and export policies per-peer (see the example in figure \ref{fig:irr_block_example}).
Usually, an AS will accept (import policy) all routes from its providers and will announce (export policy) all routes to its clients.  An AS will accept from a client only this client's routes (and the client's client routes), but in some cases, ASes are sloppy and accept all client routes, assuming they will only announce their own.  This is an example of why mining IRR data is not straight-forward.

Specifically, we search for lines of the form
\begin{verbatim}
import: from <Peer> accept <ASNs>
export: to <Peer> announce <ASNs>
\end{verbatim}
where <Peer> is the peer ASN for which the rule applies, and <ASNs> can be either a specific ASN, a comma-separated list of ASNs, an AS-SET, or the word "ANY".
As mentioned above, every AS-SET is translated to its corresponding ASNs.  While RPSL~\cite{RFC2622} is significantly richer than the expressions we ingest in our analysis, almost all IRR AUT-NUM objects are using only these simple expressions.

In both policies, ASes can be referred by using one or more ASNs, AS names, or AS-SETs.
We translate every AS reference to the corresponding ASNs.

Following Gao~\cite{Gao} and many others, we use the following "valley-free" assumptions.
A customer ASN will announce all its APs and its customer APs to its provider while accepting all the routes that its provider announces; thus, we expect the customer import expression to have 'ANY' in the  \texttt{<ASNs>} field.
Similarly, we expect a provider ASN to announce all his routes to customers while accepting only the routes for his customer's APs and its customer APs. Thus, we expect the provider export expression to have 'ANY' in the \texttt{<ASNs>} field.
Two peer ASNs will announce only their APs and their customer APs while accepting only the routes for their peer's APs and its customer APs.

Table~\ref{tab:import_export_policies} summarizes the rules used for the analysis based on the observations above. Based on these rules, we generate the tuples \{AS1, AS2, ToR\} such that AS1 is the one associated with AUT-NUM object and AS2 is extracted from the \texttt{<Peer>} field of the import-export policies.

\begin{table}[h!]
  \caption{Import-Export Policy Heuristic.}
  \centering
  \begin{tabular}{| c | c | c |}
    \hline
    \textbf{Role} & \textbf{Import} & \textbf{Export} \\ \hline
    Customer & ANY & specified ASes \\ \hline
    Provider & specified ASes & ANY \\ \hline
    Peer & specified ASes & specified ASes \\ \hline
  \end{tabular}
  \label{tab:import_export_policies}
\end{table}

Finally, there are cases where an AS announces ANY in both its import and export policies for a specific ASN. In this case, we estimate the importance of the two ASes by counting the number of their export ANY statements and taking the larger one as the provider of the specific ToR.
This heuristic was applied to 12,251 ToRs. 4,303 of the ToRs exist in the opposite direction, i.e., we can infer the ToR also from the other AUT-NUM object and use it as verification. In 91.5\% of the bi-directional cases, our heuristic agreed with the other side object.

\subsubsection{Remarks Heuristic}

In many cases (see the example in Figure \ref{fig:irr_block_example}), the AUT-NUM object contains additional {\em remark} lines that provide ToR information. The remarks usually signal the beginning of a {\em remark-block} of import/export expressions for a specific type of ToR.
We use the following keywords, which we found to be mostly used, to identify and classify the beginning of a block (the remark-block is the entity for this analysis):
\begin{enumerate}
    \item P2C - 'downstream', 'downlink', 'customer', 'client'.
    \item C2P - 'provider', 'upstream', 'uplink', 'transit'.
    \item P2P - 'peer'.
\end{enumerate}
We identify the end of a block by either of the following:
\begin{itemize}
    \item A new keyword from above.
    \item The word 'end.'
    \item The end of the AUT-NUM object.
\end{itemize}


The keyword 'peer' can be used to describe all three ToRs. Therefore, we only use it iff both P2C and C2P related keywords have already been presented previously in the same AUT-NUM object.
In addition, in many cases, the keyword 'transit' is concatenated with other keywords and can also describe all three ToRs. Therefore, we only use the 'transit' keyword when no other keywords are presented.

For each block, we generate a list of tuples \{AS1, AS2, ToR\} such that AS1 is the one associated with AUT-NUM object and AS2 is extracted from the \texttt{<Peer>} field of the import-export policies.  Later, when we discuss accuracy analysis for this heuristic, we perform the analysis per identified block.

\subsubsection{Sets Heuristic}
As mentioned above, sets in the AS-SET objects can act as an alias for multiple ASes. However, in many cases, set names provide information about ToRs, e.g., the set "AS247:AS-Customers". 
We look for common keywords and deduce the ToR for each set member using the following keywords:
%
\begin{enumerate}
    \item P2C - 'downstream', 'downlink', 'customer', 'client', 'custs'.
    \item C2P - 'provider', 'upstream', 'uplink', 'backbone'.
    \item P2P - 'peer'.
\end{enumerate}

This heuristic predicts only 26,922 links, which is significantly smaller than the other two. However, 85.3\% of its predictions (22,964 links) are entirely new since it relies on different data than the import/export and the remarks heuristics.

\subsection{Filtering Unreliable Data}

There are numerous reasons for our heuristics to fail.  For example, the word 'peer' is used to identify P2P relationships but sometimes describes BGP peering (namely, a BGP connection of any ToR).
Remarks often specify a company's name, and names may include any of the keywords. For example, the word 'provider' appears in the name of many ASes, such as "SolNet Internet Solution Provider"; and will result in false classifications for the entire remark-block.

To overcome erroneous classification, each of the three heuristics can be filtered using the fact that many AS links can be mined from both ends; we term them {\em bi-directional links}.
Filtering the data is done at the granularity of the {\em entity} that is used by each heuristic: an ASN (or AUT-NUM) for 'import-export', a remark-block for 'remarks', and an as-set for 'set'.
The import-export heuristic classifies significantly more AS links. 
Moreover, sets and remarks are typically used by large organizations. Thus most links will not be classified by these heuristics from both ends.
Therefore, to grade the reliability of our mining from each entity, we use the import-export classification of the other end of the link, if it exists. Then, we obtain the number of bi-directional links for each entity and check how many of them have ToR classification agreements between the two ends.

We use the following terms in the filtering process and its analysis:
\begin{itemize}
    \item $L(i)$ - the set of AS links per entity $i$. 
    \item $B(i)$ - the set of bi-directional AS links per entity $i$. 
    \item $A(i)$ - the set of bi-directional links with an agreement between the two ends per entity $i$.
\end{itemize}

Our first step in filtering the data is to understand how reliable is our extraction process per entity.  For this end, we define the {\em entity reliability grade}, $r(i)$, which is calculated as the portion of reliable bi-directional links per entity.  We observe that for many small ASNs, the number of bi-directional links is too small to draw a conclusion about the process reliability, thus we define a threshold, $T_b$, such that only ASNs with at least $T_b$ bi-directional link are assigned a positive reliability grade, namely 
\begin{equation}
  r(i) = \begin{cases}
        |A(i)|/|B(i)| & \text{$|B(i)| \ge T_b$,}
        \\
        0 & \textrm{otherwise}.
        \end{cases}
 \end{equation}

To determine the ToR of an AS link, we use the end (if the ToR can be deduced from both ends) with the higher entity reliability grade, and the ToR reliability inherits this end's reliability. Namely, $r(i,j)=\max\{r(i), r(j)\}$. In case that a ToR can be deduced from only one entity, the ToR inherits this entity's reliability grade.

We want to determine the conditions where our results can be trusted with high confidence. Obviously, if we filter out entities with low-reliability grades and a small number of bi-direction links (due to the incomplete statistics), we will increase the trustworthiness of the results.  However, the stricter we are, the number of links we can classify will decrease. To measure this trade-off, we first define two measures:
the agreement ratio, $R$, which is calculated by a portion of bi-directional links where both ends agree on their ToR:
\begin{equation}\label{eq:r}
    R = |\cup_{i}A(i)|/|\cup_{i}B(i)|,~~\forall i : r(i) \ge T_r,
\end{equation}
where $T_r$ is the minimum entity reliability grade which is required in order to consider this entity in the analysis,
and the link coverage which is calculated by
\begin{equation}
    C = |\cup_{i}L(i)|/|\cup_{j}L(j)|,~~\forall i : r(i) \ge T_r, \forall j.
\end{equation}

\begin{figure}[h!]
   \begin{subfigure}[h]{0.45\textwidth}
    \centering
    \includegraphics[width=\linewidth]{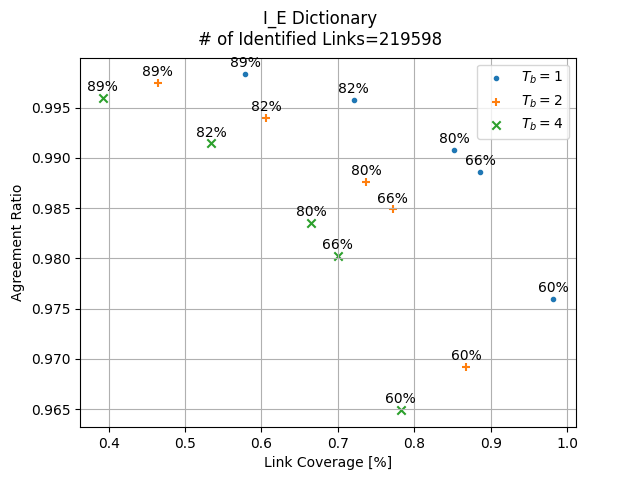}
    \caption{Agreements ratio vs. link coverage}\label{fig:ie_params:a}
    \end{subfigure}
    \begin{subfigure}[h]{0.45\textwidth}
    \centering
    \includegraphics[width=\linewidth]{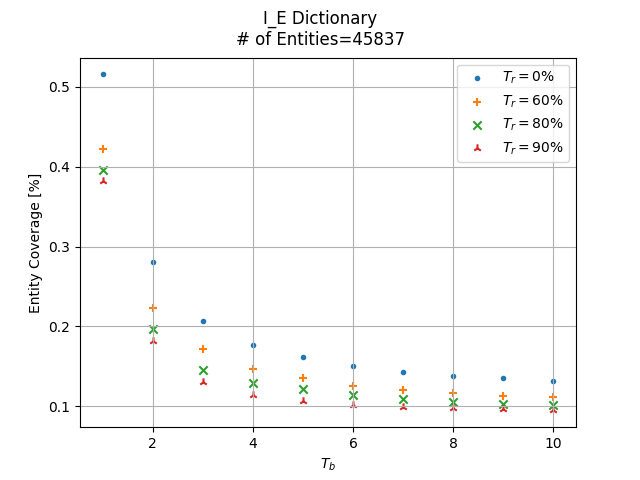}
    \caption{Entity coverage vs. $T_b$}\label{fig:ie_params:b}
    \end{subfigure}
    \begin{subfigure}[h]{0.45\textwidth}
    \centering
    \includegraphics[width=\linewidth]{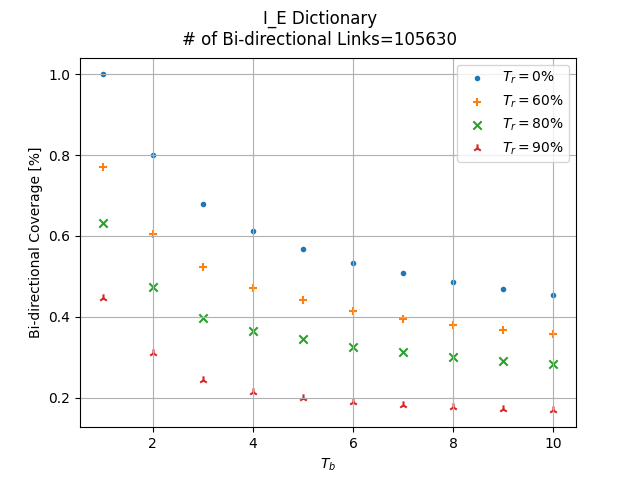}
    \caption{Bi-directional link coverage vs. $T_b$}\label{fig:ie_params:c}
    \end{subfigure}
    
    \caption{Import-Export Heuristic Parameters Exploration}\label{fig:ie_params}
\end{figure}

\begin{figure}[h!]
   \begin{subfigure}[h]{0.45\textwidth}
    \centering
    \includegraphics[width=\linewidth]{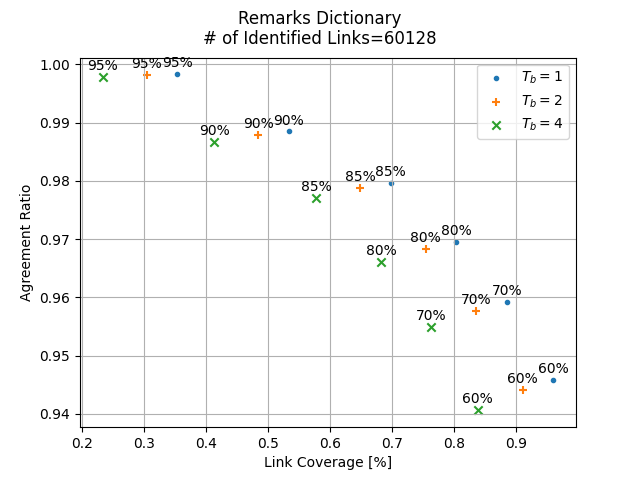}
    \caption{Agreements ratio vs. link coverage}\label{fig:remarks_params:a}
    \end{subfigure}
    \begin{subfigure}[h]{0.45\textwidth}
    \centering
    \includegraphics[width=\linewidth]{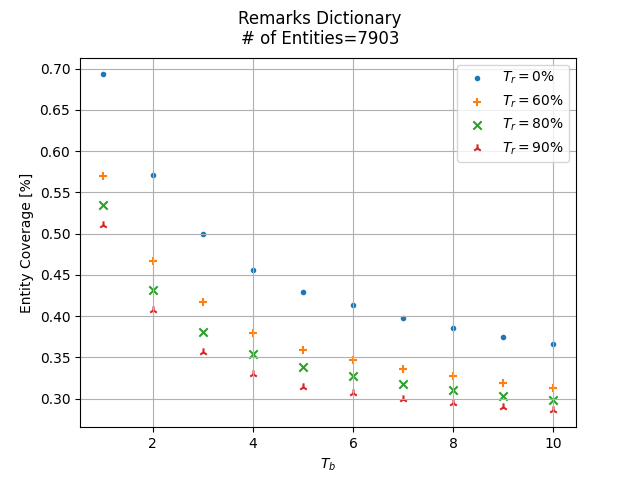}
    \caption{Entity coverage vs.\ $T_b$}\label{fig:remarks_params:b}
    \end{subfigure} 
    \begin{subfigure}[h]{0.45\textwidth}
    \centering
    \includegraphics[width=\linewidth]{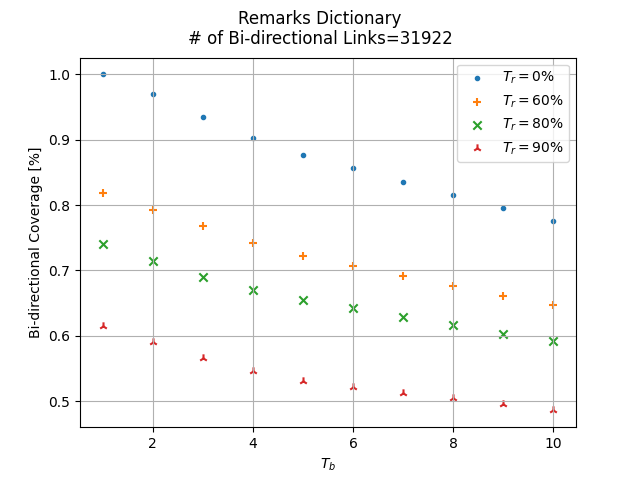}
    \caption{Bi-directional link coverage vs.\ $T_b$}\label{fig:remarks_params:c}
    \end{subfigure}
    
    \caption{Remarks Heuristic Parameters Exploration}\label{fig:remarks_params}
\end{figure}

\begin{figure}[h!]
   \begin{subfigure}[h]{0.45\textwidth}
    \centering
    \includegraphics[width=\linewidth]{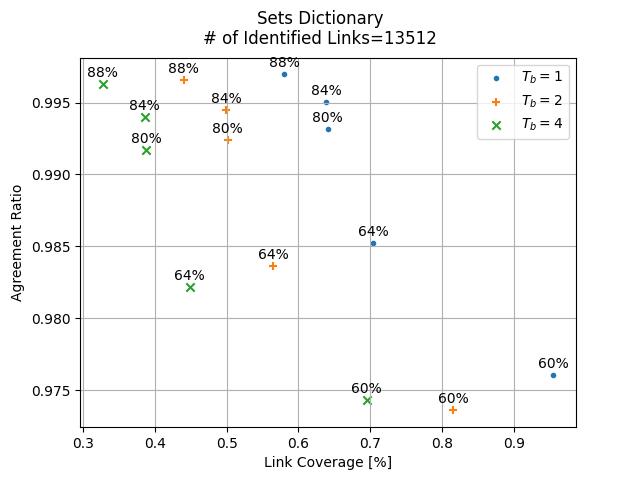}
    \caption{Agreements ratio vs.\ link coverage}\label{fig:sets_params:a}
    \end{subfigure}
    \begin{subfigure}[h]{0.45\textwidth}
    \centering
    \includegraphics[width=\linewidth]{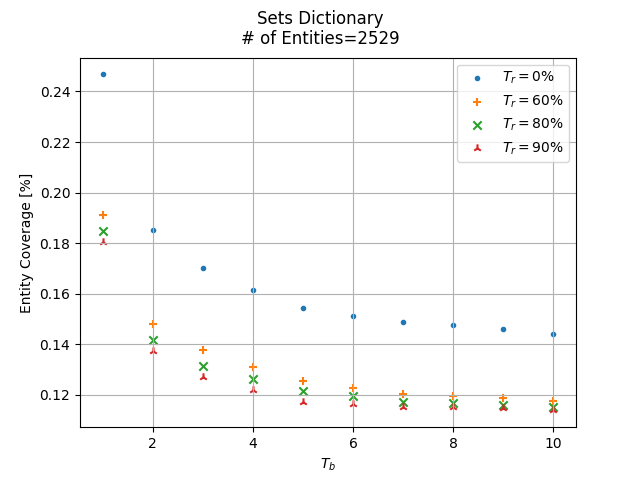}
    \caption{Entity coverage vs.\ $T_b$}\label{fig:sets_params:b}
    \end{subfigure} 
    \begin{subfigure}[h]{0.45\textwidth}
    \centering
    \includegraphics[width=\linewidth]{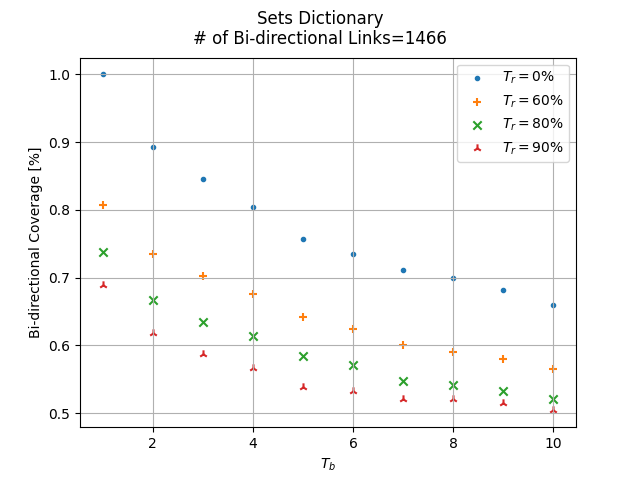}
    \caption{Bi-directional link coverage vs. $T_b$}\label{fig:sets_params:c}
    \end{subfigure}
    
    \caption{Sets Heuristic Parameters Exploration}\label{fig:sets_params}
\end{figure}

\subsection{Siblings Inference}
Previous works~\cite{CAIDA07,coretoleaf_19}  offered to infer siblings based on their organization name. In this work, we extend this idea and show that siblings can be inferred from multiple fields.
Our siblings' inference is based on matches in multiple administrative fields of the AUT-NUM object: 
\begin{enumerate}
    \item {\em "changed"} - specifies the email address of the last person that changed the AUT-NUM block. We extract the {\em domain} of the email address.
    \item {\em "org"} - specifies the organization that is registered with the AS.
    \item {\em "mnt-by"} - specifies the responsible party for the AS's maintenance.
    \item {\em "admin"} - specifies the administrator of the AS.
    \item {\em "tech"} - specifies the technical contact for the AS.
    \item {\em "notify"} - specifies the contact's email for the AS.
    \item {\em "AS-name"} - specifies the name of the AS.
\end{enumerate}
In addition, we use the {\em "mnt-by"} field for AS-SET objects, which specifies the responsible party for the AS-SET's maintenance.

As mentioned before, we first filter the objects and take only the latest updated version of each object.  In addition, we filter out object fields as follows:
\begin{enumerate}
    \item The top 20 used domain names (such as 'Gmail' and 'Hotmail').
    \item Any field that includes the word 'dum,' since this usually means that the field is a dummy field with no real information.
    \item AS-names field with the 'unspecified' value.
\end{enumerate}

Some fields ("admin," "changed," "mnt-by," "org") can mention maintenance companies that handle a large number of ASes with no real connection between them. Usually, maintenance companies appear more times than we expect when looking for siblings.

\section{Results}\label{sec:results}

In this section we present the results we obtained from running our heuristics.  In the next section we will evaluate their accuracy by comparing them to previous results and manual ground truth datasets.

\subsection{Filtering Results}
Figures~\ref{fig:ie_params:a}, \ref{fig:remarks_params:a}, \ref{fig:sets_params:a} depict the agreement ratio $R$ versus the link coverage $C$ as a function of the two thresholds; $T_r$ (written above each mark) and $T_b$ (the mark type and color). The most notable observation from the figures is that the agreement ratio is very high for all three heuristics, above 94\%, and for the heuristic that produces the largest dataset (Import/Export), it is above 96\%. Namely, for all the tested parameters, the data coherency in the IRR is quite small, which we believe indicates a small classification error. 

Surprisingly, a lax requirement from the minimum number of bi-directional links, $T_b$, has little influence on the agreement ratio: almost no decline in the agreement ratio when lowering this threshold from 4 to 1 (results for $T_b=5$ are also almost identical to $T_b=4$, thus removed from the plots), and a decline of about 1\% when setting the threshold at 1.  The gain in coverage is substantial: up to 10 percentage points gain when lowering the threshold from 4 to 2, and up to additional 10 percentage points when reducing the threshold to 1.

Figures~\ref{fig:ie_params:b}, \ref{fig:remarks_params:b}, \ref{fig:sets_params:b} depict how the thresholds influence the number of entities (ASNs and AS-SETs) that remain after filtering. We first notice that the initial number of entities that are available for the I/E heuristic is 45,837 but is quite small for the other two heuristics: 7,903 for Remarks and 2,529 AS-SETs.  It is also clear that the requirement for two-directional links reduces the number of entities, significantly.

Figures~\ref{fig:ie_params:c}, \ref{fig:remarks_params:c}, \ref{fig:sets_params:c} depict how the thresholds influence the number of bi-directional links.  Note that this is different than the number of discovered links that is used in Figures~\ref{fig:ie_params:a}, \ref{fig:remarks_params:a}, \ref{fig:sets_params:a}.
These 6 figures show that while we lose a significant number of entities in the filtering, the percentage of lost bi-directional links is more modest.
Figures~\ref{fig:ie_params:a}, \ref{fig:remarks_params:a}, \ref{fig:sets_params:a} show that the number of links in general is only slightly reduced since the ASNs that are left from the filtering process cover also links that were filtered out from other entities.

Thus, taking 97.5\% agreement ratio as a reasonable working point, we can use $T_b=1$ and $T_r=60\%$ for the I/E and Sets heuristics and $T_r=80\%$ for the Remarks heuristic.

\subsection{Filtering Consistency} 

Using the thresholds from the previous subsection, we examine
the consistency of results between the three heuristics by comparing the import/export heuristic with the other two heuristics (See Tables~\ref{tab:remarks_ie_compare} and \ref{tab:sets_ie_comapare}). The Remarks and Sets heuristics will not be compared since the overlap between their predictions is minimal (88 links).

Tables~\ref{tab:remarks_ie_compare} and \ref{tab:sets_ie_comapare} have four parts: the upper left part is a 'confusion matrix', which compares the I/E labeling with the Remarks or Sets labeling.  For this comparison, we duplicate each ToR, such that if a link (AS1, AS2) exists, we also generate its reverse link (AS2, AS1) with the proper ToR. 
The bottom row shows the precision for each ToR, the right column shows the recall for each ToR, and the bottom-right cell holds the overall agreement.  Note that for convenience, the terms here assume that the I/E data is the ground truth. 

The overlap between the I/E and Remarks classifications is quite large, 46,719 ToRs, which are 87.8\% of the classifications that were left in the Remarks predictions after the filtering.
The agreement between the two heuristics is quite high: 97.5\%, which attests to the reliability of the post-filtering classification.

The overlap between the I/E and Set classifications is quite small, only 1,973 ToRs, which are less than 15.0\% of the classifications that were left in the Sets predictions after the filtering.
The agreement between the two heuristics is very high: 98.7\%, which again attests to the consistency of our filtering method.

\begin{table}[h!]
  \caption{Remarks vs.\ Import/Export Comparison.} \label{tab:remarks_ie_compare}
  \centering
    \begin{tabular}{|l|c|c|c||c|}\hline
     & \thead{Remarks \\ P2P} & \thead{Remarks \\ P2C} & \thead{Remarks \\ C2P} & \thead{Recall} \\ \hline
    \thead{I/E  P2P} & 42,474 & 772 & 772 & 96.7\% \\ \hline
    \thead{I/E  P2C} & 322 & 24,297 & 91 & 98.3\% \\ \hline
    \thead{I/E  C2P} & 322 & 91 & 24,297 & 98.3\% \\ \hline\hline
    \thead{Precision} & 98.5\% & 96.6\% & 96.6\% & 97.5\% \\ \hline
\end{tabular}
\end{table}

\begin{table}[h!]
  \caption{Sets vs.\ Import/Export Comparison.}
  \label{tab:sets_ie_comapare}
  \centering
    \begin{tabular}{|l|c|c|c||c|}\hline
     & \thead{Sets \\ P2P} & \thead{Sets \\ P2C} & \thead{Sets \\ C2P} & \thead{Recall} \\ \hline
    \thead{I/E  P2P} & 304 & 17 & 17 & 90.0\% \\ \hline
    \thead{I/E  P2C} & 7 & 1,796 & 1 & 99.6\% \\ \hline
    \thead{I/E  C2P} & 7 & 1 & 1,796 & 99.6\% \\ \hline\hline
    \thead{Precision} & 95.6\% & 99.0\% & 99.0\% & 98.7\% \\ \hline
\end{tabular}
\end{table}

\subsection{Heuristics Union}\label{sec:heuristics_union}
Given the three sets of ToR classifications, one per heuristic, we want to generate a single classification and reliability score. 


We use the following algorithm for the final classification:
\begin{enumerate}
    \item Majority vote: if two or more heuristics classify a link with the same ToR - we set the reliability score to 1 and choose this ToR.
    \item If none of the heuristics agree, the final classification is the ToR with the highest reliability score. In this case, the final reliability score is the reliability score of that ToR.
    \item If none of the heuristics agree, and two or more of them have the same reliability score, the ToR from the entity with the highest number of bi-directional links ($|B(i)|$) is taken with its related reliability score.
    \item If all the reliability scores are the same, with the same number of bidirectional links, the I/E heuristics classifications is selected.
\end{enumerate}
Note that not every link is classified by the three heuristics. The majority of the links are classified by only one or two heuristics, such that 75\% of the links (291,474 links) are classified by only one heuristic, almost 25\% of the links (97,244 links) are classified by two heuristics, and only 76 links are classified by all three heuristics.
There were 54 AS links where both heuristics have the same reliability score (1.0) and the same small number of bidirectional links. For these cases, we selected the I/E heuristics classifications.

\section{Evaluation}\label{sec:eval}

\subsection{Datasets}\label{sec:dataset}
To evaluate the accuracy of our results and compare them with previously published methods, we use two collections of manually inspected ToRs. The datasets were collected between January 2018 and December 2021. The ToRs are extracted from the proprietary BGProtect ToR database, which is populated using multiple algorithms that are based on both IRR and BGP data.  The BGProtect database also contains a limited number of entries that were manually entered, usually where algorithmic methods could not generate a ToR or when a route with valley-free violation~\cite{Gao} was examined by an expert analyst who had found an error in a ToR.  

Analysts manually updates an entry only if they find conclusive evidence for the ToR.
Manual entry of a ToR is usually documented with a short justification stating the strongest evidence used to determine the ToR.  Routing data is always available to the analyst since the main reason for a manual examination of a ToR is a missing ToR in a monitored route or a routing anomaly in a route flagged by the system. In addition, IRR data, AS size and role are also considered. We obtained two groups of ToRs whose justification was made clear:
\begin{enumerate}
    \item {\em Manual IRR}: a collection of 412 ToRs (31 P2P, 381 P2C) that were manually inserted by analysts where the main justification was examining the IRR (usually there is additional supporting evidence such as routing data or type of ASes to support such an insertion).  This dataset helps validate that the automatic process we suggest is coherent with a human understanding of the data.
    \item {\em Manual Routing}: 288 manual entries (10 P2P, 278 P2C), where the ToR justification was routing (either BGP or traceroute, usually from distinct time epochs) and IRR was not mentioned.  These entries represent cases where the respected IRR entries are inconclusive or missing, thus we do not expect our algorithm to perform well on this dataset.
\end{enumerate}
We will also use the unification of both datasets and term it {\em Unified Manual} dataset.

To ensure that the manual datasets are not biased, we examined the roughly 1400 end-points that participate in the ToRs, and found 875 unique ASNs in both datasets. The most frequent ASNs in the dataset are AS15830 which appears 20 times, AS12389 which appears 14 times, and AS204136 which appears 13 times.  The ASN at the 10th place appears only 9 times.  There is little representation to tier-1 providers since their ToRs are usually quite obvious, none of them appear among the top 20 ASNs. These ASNs are registered in 92 nations: 95 ASNs from Russia, 92 ASNs from the US, and 40 ASNs from Brazil and Iran.  Note that there is some bias towards nations with problematic ToRs like Brazil and Iran (see the geo-filtering in \cite{AP2Vec}).





\subsection{Evaluation Criteria}\label{sec:eval_criteria}
To assess the performance of our model, we use accuracy as the primary evaluation metric, defined as the proportion of correctly classified examples out of the total predictions made. Formally, for a multiclass classification task, accuracy is given by:  

\begin{equation}
    Accuracy = \frac{\sum_{i\in classes}TP_{i}}{\sum_{i\in classes}(TP_{i} + FP_{i})},
\end{equation}  

where \(TP_i\) and \(FP_i\) represent the true positives and false positives for class \(i\), respectively.  

Additionally, we compute recall and precision for each class. Recall (\(Rc\)) measures the proportion of correctly identified positive instances and is defined as:  

\begin{equation}
    Rc = \frac{TP}{TP + FN},
\end{equation}  

where \(FN\) denotes false negatives.  

Similarly, precision (\(PR\)) quantifies the fraction of correctly predicted positive instances and is given by:  

\begin{equation}
    PR = \frac{TP}{TP + FP}.
\end{equation}


\subsection{Initial Heuristics Results}
Using data from 45,837 ASes, the I/E heuristic was able to classify 327,227 ToRs between 66,834 ASes.  105,630 of our classifications are 2-sided (namely, they classify 52,815 links), while the remaining 221,597 are 1-sided classifications. Out of the 105,630 2-sided classifications, there is an agreement of 82.3\%.

With the remarks heuristic, we could analyze a significantly smaller number of entities, only 3,288.
The remarks heuristic was able to classify 81,563 ToRs between 21,149 ASes. 31,922 of our classifications are 2-sided (namely, they classify 15,961 links), while the remaining 49,614 are 1-sided classifications. Out of the 31,922 2-sided classifications, there is an agreement ratio of 80.48\%.

The sets heuristic produced even fewer data.
Using 4,940 entities, the sets heuristic was able to classify 45,816 ToRs between 21,779 ASes. 1,466 of our classifications are 2-sided (namely, they classify 733 links), while the remaining 44,350 are 1-sided classifications. Out of the 1,466 2-sided classifications, there is an agreement ratio of 82.7\%.

The overlap between the I/E results and the Unified Manual dataset is 539 ToRs. For the Remarks dataset, it is 236 ToRs, and for the SETs, it is 12 ToRs.

Tables \ref{tab:initial_ie_results} and \ref{tab:initial_remarks_results} have four parts: the upper left part is a confusion matrix, which compares our labeling with the {\em unified manual} labeling. The bottom row shows the precision for each ToR, the right column shows the recall for each ToR, and the bottom-right cell holds the overall accuracy.  

For all the heuristics, the precision of  P2P ToRs is the lowest.  This is expected since P2P connections are the hardest to deduct. The accuracy of both the I/E and remarks heuristics is close to 89\%, for sets heuristic, the overlap is only 12 ToRs, and only 1 is wrongly classified.

\begin{table}[h!]
  \caption{Import/Export heuristic initial results.}
  \centering
    \begin{tabular}{|l|c|c|c||c|}\hline 
     & \thead{Predicted \\ P2P} & \thead{Predicted \\ P2C} & \thead{Predicted \\ C2P} & \thead{Recall} \\ \hline
    \thead{Manual P2P} & 39 & 0 & 0 & 100.0\% \\ \hline
    \thead{Manual P2C} & 17 & 183 & 29 & 79.9\% \\ \hline
    \thead{Manual C2P} & 9 & 6 & 256 & 94.5\% \\ \hline\hline
    \thead{Precision} & 60.0\% & 96.8\% & 89.8\% & 88.7\% \\ \hline
\end{tabular}
\label{tab:initial_ie_results}
\end{table}


\begin{table}[hb!]
  \caption{Remarks heuristic initial results.}
  \centering
    \begin{tabular}{|l|c|c|c||c|}\hline
     & \thead{Predicted \\ P2P} & \thead{Predicted \\ P2C} & \thead{Predicted \\ C2P} & \thead{Recall} \\ \hline
    \thead{Manual P2P} & 15 & 4 & 1 & 75.0\% \\ \hline
    \thead{Manual P2C} & 6 & 122 & 8 & 89.7\% \\ \hline
    \thead{Manual C2P} & 4 & 5 & 71 & 88.8\% \\ \hline\hline
    \thead{Precision} & 60.0\% & 93.1\% & 88.8\% & 88.1\% \\ \hline
\end{tabular}
\label{tab:initial_remarks_results}
\end{table}

\subsection{Filtered Heuristics Results}



After filtering, we succeeded in classifying 178,907 ToRs between 39,629 ASes of the I/E heuristic, 50,735 ToRs between 18,478 entities of the remarks heuristic, and 13,461 ToRs between 9,747 entities of the sets heuristic.
There is an overlap of 582, 322, and 10 ToRs between the {\em unified manual} dataset and the I/E, Remarks and Sets classifications, respectively.
Note, that the overlap now is greater than when each heuristic was used individually, since before we examined only bi-directional links and here we add links based on 1 trusted end-point.

After the filtering (see Tables \ref{tab:filtered_ie_results} and \ref{tab:filtered_remarks_results})
the accuracy of our classifications increased from 88.7\% and 88.1\% to 93.1\% and 95.5\%, respectively. Remember that the agreement between the 3 heuristics is also around 98\% (see Tables~\ref{tab:remarks_ie_compare} and \ref{tab:sets_ie_comapare}).
This means that our filtering successfully removes most of the questionably classified ToRs and produces a relatively large number of highly reliable classifications.
Notice that for this comparison (and the one in Table~\ref{tab:unified_results}), we duplicate each ToR, such that if a link (AS1, AS2) exists, we also generate its reverse link (AS2, AS1) with the proper ToR. 
 For the set heuristic, there were only 10 overlapping ToRs, 8 of them were correctly classified.

\begin{table}[h!]
  \caption{Filtered Import/Export heuristic results.}
  \label{tab:filtered_ie_results}
  \centering
    \begin{tabular}{|l|c|c|c||c|}\hline
     & \thead{Predicted \\ P2P} & \thead{Predicted \\ P2C} & \thead{Predicted \\ C2P} & \thead{Recall} \\ \hline
    \thead{Manual P2P} & 56 & 0 & 0 & 100.0\% \\ \hline
    \thead{Manual P2C} & 10 & 243 & 10 & 92.4\% \\ \hline
    \thead{Manual C2P} & 10 & 10 & 243 & 92.4\% \\ \hline\hline
    \thead{Precision} & 73.7\% & 96.0\% & 96.0\% & 93.1\% \\ \hline
\end{tabular}
\end{table}


\begin{table}[h!]
  \caption{Filtered Remarks heuristic results.}
  \label{tab:filtered_remarks_results}
  \centering
    \begin{tabular}{|l|c|c|c||c|}\hline
     & \thead{Predicted \\ P2P} & \thead{Predicted \\ P2C} & \thead{Predicted \\ C2P} & \thead{Recall} \\ \hline
    \thead{Manual P2P} & 28 & 0 & 0 & 100.0\% \\ \hline
    \thead{Manual P2C} & 6 & 140 & 1 & 95.2\% \\ \hline
    \thead{Manual C2P} & 6 & 1 & 140 & 95.2\% \\ \hline\hline
    \thead{Precision} & 70.0\% & 99.3\% & 99.3\% & 95.7\% \\ \hline
\end{tabular}
\end{table}

\subsection{Unified Results}\label{sec:unified_results}
Overall, we were able to classify 194,397 ToRs between 70,914 ASes.
Table \ref{tab:unified_results} summarizes the comparison with the unified manual dataset.  For the comparison, we duplicate all links, namely the link AS1-AS2 was duplicated for the link AS2-AS1.
The overlap with the ground truth had 30 P2P links, all of them were correctly classified. 253 of the 272 C2P links (93.0\%) were classified correctly.
Using the previously defined confidence measure, we achieved 103,653 classifications with 100\% reliability and 148,847 classifications above 90\% reliability. 


\begin{table}[h!]
  \caption{Unified results.}
  \label{tab:unified_results}
  \centering
    \begin{tabular}{|l|c|c|c||c|}\hline
     & \thead{Predicted \\ P2P} & \thead{Predicted \\ P2C} & \thead{Predicted \\ C2P} & \thead{Recall} \\ \hline
    \thead{Manual P2P} & 60 & 0 & 0 & 100.0\% \\ \hline
    \thead{Manual P2C} & 11 & 253 & 8 & 93.0\% \\ \hline
    \thead{Manual C2P} & 11 & 8 & 253 & 93.0\% \\ \hline\hline
    \thead{Precision} & 73.2\% & 96.9\% & 96.9\% & 93.7\% \\ \hline
\end{tabular}
\end{table}

\subsection{Siblings Results}
\begin{figure}[h!]
    \centering
    \includegraphics[width=0.8\linewidth]{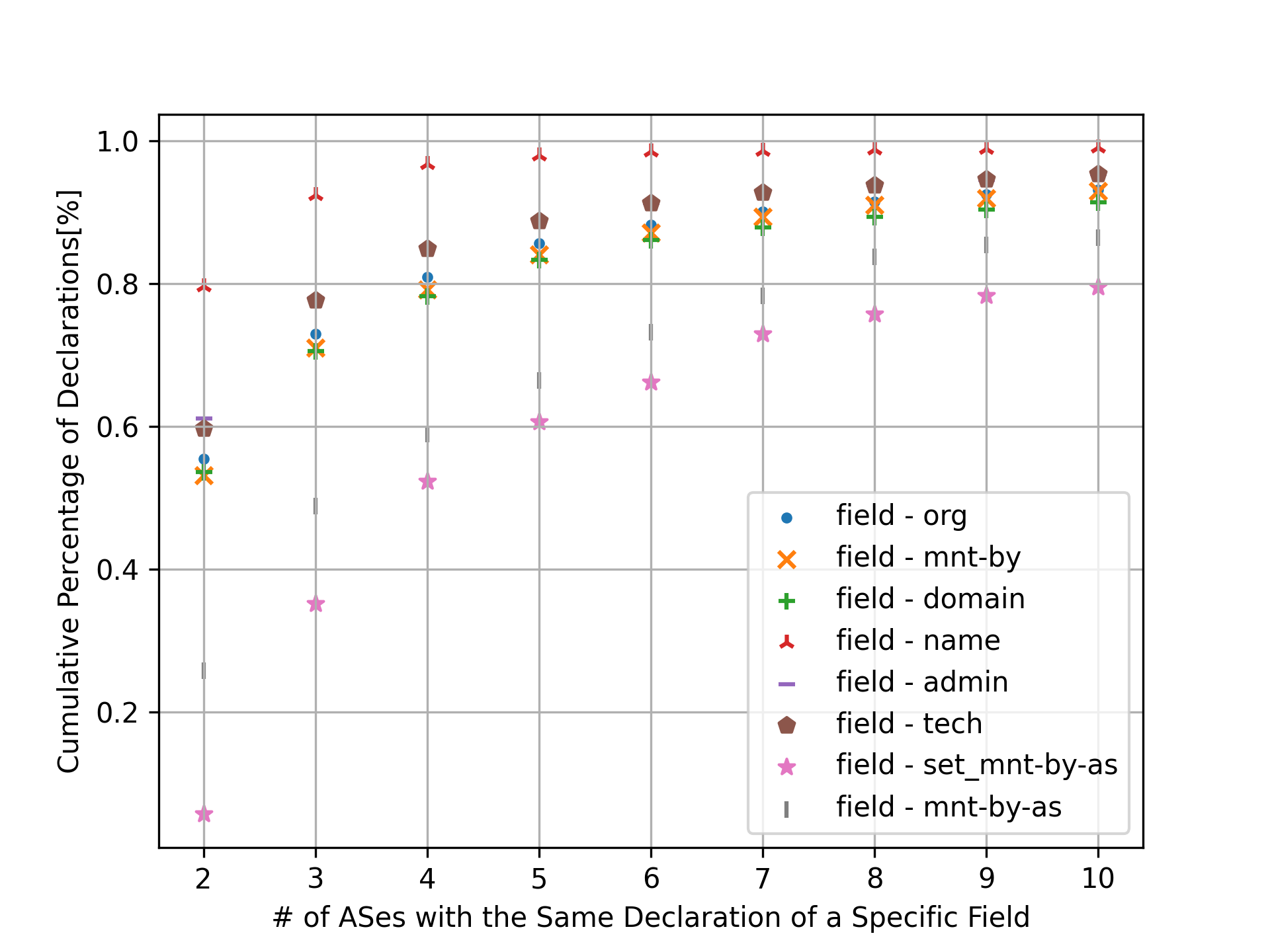}
   \caption{CDFs of the number of ASes with the same declaration for each field used for siblings inference.}\label{fig:siblings_cdfs}
\end{figure}

Figure~\ref{fig:siblings_cdfs} presents the CDFs of the number of ASes with the same value of a specific field.
Figure~\ref{fig:siblings_cdfs} shows that most of the matches (over 80\%) for all the fields only appear in a few entries (5 or less). Thus we can safely set a small threshold on this number.
To set the threshold, we use the observation that matching names in the AS-name field can rarely be coincidental.  Indeed, Figure~\ref{fig:siblings_cdfs} shows that almost all the matches for the 'name' field are seen in 2-5 ASNs, thus for all the fields, we chose to include only entries that match less than 6 times.


We compare our results with two datasets: the ProbLink \cite{coretoleaf_19} dataset from April 2021 that contains 2,834 siblings, which uses the CAIDA’s AS-to-organization mapping dataset to infer links between ASes that are operated by the same organization as sibling relationships; and with a proprietary dataset of BGProtect~\cite{BGPROTECT} that has 2,990 siblings. The CAIDA database~\cite{caida_asrelationship} does not contain siblings.
Table~\ref{tab:siblings_resuts} shows a comparison with ProbLink for each of the fields in the AUT-NUM object. We present both the overall findings of the specific field and the links that were uniquely discovered only by this field.  The discoveries columns count the number of AS links that can be deducted from a field, e.g., if there are four ASNs with the same field entry, the number of discovered links is $(4\times 3)/2=6$.   

Our heuristic discovered 24,743 siblings, 2,020 of these ToRs appear in the ProbLink dataset, but only 794 appear as siblings (39.3\%).  The overlap with the BGProtect dataset has 1,029 ToRs, 739 appear as siblings (71.8\%). The overall matching, as well as the matching of each field, is higher for the non-unique discoveries.  The overall matching with ProbLink for non-unique discoveries is 49.9\%, while for the unique discoveries, it is only 26.6\%. This can be due to a combination of two reasons:  the unique identification of our heuristic introduces errors, and the uniquely identified siblings are harder to identify by ProbLink. The high matching with BGProtect data (71.8\%) indicates that the latter is the more dominant cause.

It is important to note that a mismatch of a sibling ToR may not be an error.  Some networks organize their ASNs by using one network as a 'backbone' that connects the other ASNs to the rest of the Internet. In other cases, networks may have ISP activity and cloud or services activities using different ASNs. Thus, while we will classify them as siblings since they belong to the same organization, databases based on routing behavior (like ProbLink and BGProtect) will deduce a P2C relationship, which is acceptable.  
For example, AS2686 is AT\&T Global Network Services, which is using the main AT\&T carrier network, AS7018, as an upstream. Indeed, in the CAIDA database~\cite{caida_asrelationship}, which is using routing data, the two appear as C2P, while our algorithm and BGProtect classify them as siblings. 

A possible reason for the significant difference between our sibling dataset size and the other datasets is that ASNs that belong to the same organization, and are thus siblings, may not be connected directly.  As a result, they do not appear on routing based datasets. Their sibling relationship is theoretical.  However, adding redundant sibling links between unconnected ASNs have no negative results for the main reason ToRs are deduced, which is the ability to identify hijacked routes (see Introduction).


\begin{table*}[h!]
\scriptsize
  \caption{Siblings results - comparison with {\em ProbLink}~\cite{coretoleaf_19}.} \label{tab:siblings_resuts}
  \centering
    \begin{tabular}{|l|c|c|c|c|c|c|}\hline
    \multirow{3}{*}{\thead{Field}} & \multicolumn{3}{|c|}{\thead{Overall}} & \multicolumn{3}{|c|}{\thead{Unique}} \\ \cline{2-7}
    & \thead{Discoveries} & \thead{Overlap\\ with ProbLink} & \thead{Match\\ with ProbLink}  & \thead{Discoveries} & \thead{Overlap\\ with ProbLink} & \thead{Match\\ with ProbLink}\\ \hline
    \thead{org} & 12,573 & 945 & 472 (49.9\%) & 5,473 & 321 & 92 (28.7\%)  \\ \hline
    \thead{mnt-by} & 11,076 & 1,216 & 475 (39.1\%) & 3,396 & 316 & 54 (17.1\%) \\ \hline
    \thead{domain} & 4,535 & 658 & 226 (34.3\%) & 1,692 & 149 & 30 (20.1\%) \\ \hline
    \thead{name} & 3,649 & 153 & 123 (80.4\%) & 2,177 & 62 & 46 (74.2\%) \\ \hline
    \thead{admin} & 4,564 & 418 & 235 (56.2\%) & 286 & 19 & 8 (42.1\%) \\ \hline
    \thead{tech} & 5,039 & 470 & 253 (53.8\%) & 498 & 32 & 9 (28.1\%)\\ \hline
    \thead{mnt-by-as} & 678 & 86 & 33 (38.4\%) & 138 & 21 & 6 (28.6\%)\\ \hline\hline
    \thead{Total} & 24,743 & 2,020 & 794 (39.3\%) & 13,660 & 920 & 245 (26.6\%)\\ \hline
\end{tabular}
\end{table*}

\begin{table*}
\scriptsize
  \caption{Evaluation of our results and other methods by using the {\em manual IRR} labels.} \label{tab:man_eval_IRR}
  \centering
    \begin{tabular}{|l|*{8}{c|}}\hline
     & \thead{Unfiltered\\IRR} & \thead{Filtered\\IRR} & \thead{CAIDA} \cite{caida_asrank} & \thead{ProbLink} \cite{coretoleaf_19} & \thead{GAO} \cite{Gao} & \thead{SARK} \cite{SARK} & \thead{RUAN} \cite{ruan} & \thead{ND-ToR} \cite{NDToR}\\ \hline
    \thead{Overlap} & 336 & 260 & 230 & 174 & 205 & 217  & 214 & 220 \\ \hline
    \thead{P2P \\ Agreement} & \thead{28/28 \\ (100.0\%)} & \thead{28/28 \\ (100.0\%)} & \thead{14/16 \\ (87.5\%)} & \thead{15/17 \\ (88.2\%)} & \thead{0/8 \\ (0.0\%)} & \thead{1/9 \\ (11.1\%)}  & \thead{4/10 \\ (40.0\%)} & \thead{3/10 \\ (30.0\%)} \\ \hline
    \thead{P2C/C2P \\ Agreement} & \thead{283/308 \\ (91.9\%)} & \thead{220/232 \\ (94.8\%)} & \thead{193/214 \\ (90.2\%)} & \thead{132/157 \\ (84.1\%)} & \thead{193/197 \\ (98.0\%)} & \thead{185/208 \\ (90.0\%)} & \thead{202/204 \\ (98.0\%)}  & \thead{204/210 \\ (97.1\%)} \\ \hline 
    \thead{Overall \\ Agreement} & 92.6\% & 95.4\% & 90.0\% & 84.5\% & 94.2\% & 85.7\%  & 95.3\% & 93.9\% \\ \hline
\end{tabular}
\end{table*}

\begin{table*}[h!]
\scriptsize
  \caption{Evaluation of our results and other methods by using the {\em manual routing} labels.} \label{tab:man_eval_route}
  \centering
    \begin{tabular}{|l|*{8}{c|}}\hline
     & \thead{Unfiltered\\IRR} & \thead{Filtered\\IRR} & \thead{CAIDA} \cite{caida_asrank} & \thead{ProbLink} \cite{coretoleaf_19} & \thead{GAO} \cite{Gao} & \thead{SARK} \cite{SARK} & \thead{RUAN} \cite{ruan} & \thead{ND-ToR} \cite{NDToR}\\ \hline
    \thead{Overlap} & 55 & 42 & 220 & 148 & 199 & 212  & 207 & 213 \\ \hline
    \thead{P2P \\ Agreement} & \thead{2/2 \\ (100.0\%)} & \thead{2/2 \\ (100.0\%)} & \thead{3/5 \\ (60.0\%)} & \thead{1/3 \\ (33.3\%)} & \thead{0/3 \\ (0.0\%)} & \thead{0/3 \\ (0.0\%)}  & \thead{1/3 \\ (33.3\%)} & \thead{0/3 \\ (33.3\%)} \\ \hline
    \thead{P2C/C2P \\ Agreement} & \thead{40/53 \\ (75.5\%)} & \thead{33/40 \\ (82.5\%)} & \thead{194/215 \\ (90.2\%)} & \thead{129/145 \\ (89.0\%)} & \thead{194/196 \\ (99.0\%)} & \thead{196/209 \\ (93.8\%)} & \thead{202/204 \\ (99.0\%)}  & \thead{205/210 \\ (97.6\%)} \\ \hline 
    \thead{Overall \\ Agreement} & 76.4\% & 83.3\% & 89.6\% & 87.8\% & 97.5\% & 82.5\%  & 98.1\% & 96.2\% \\ \hline
\end{tabular}
\end{table*}

\begin{figure}[h!]
  \begin{subfigure}[h]{0.95\linewidth}
    \centering
    \includegraphics[width=0.75\linewidth]{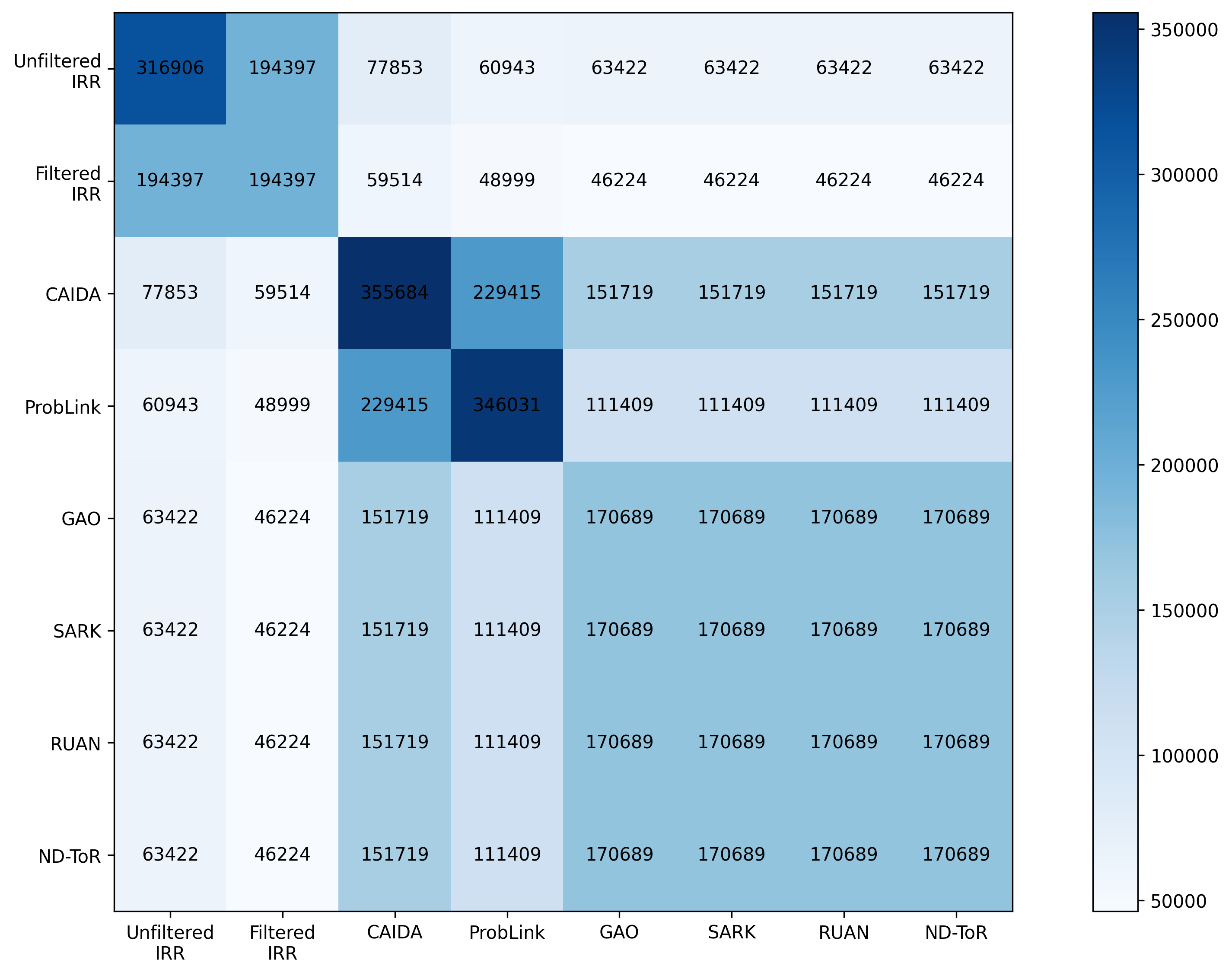}
    \caption{The overlap between different ToR classifications methods.}\label{fig:methods_overlap}
    \end{subfigure}
    \begin{subfigure}[h]{0.95\linewidth}
    \centering
    \includegraphics[width=0.75\linewidth]{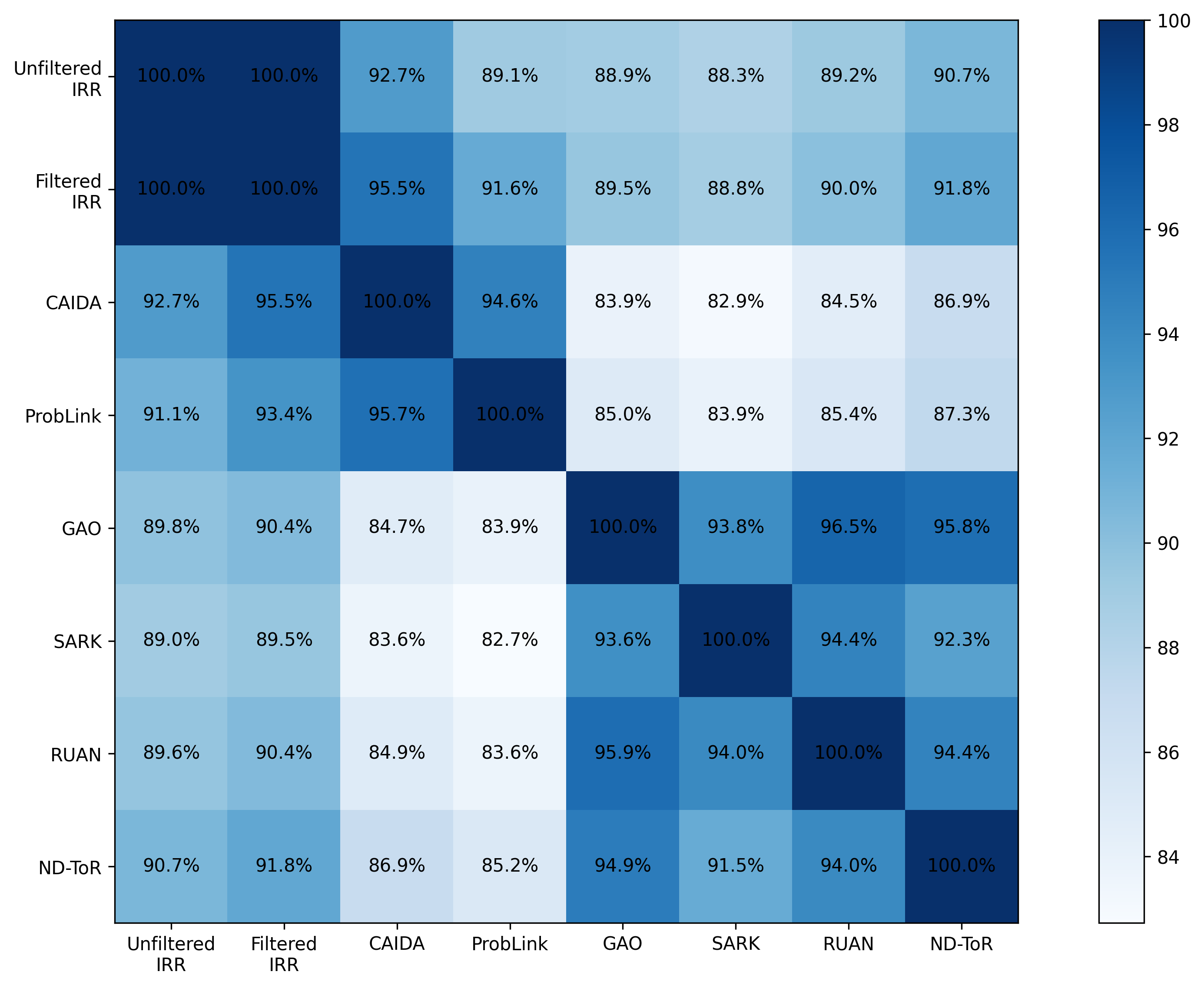}
    \caption{The agreement between different ToR classifications methods.}\label{fig:methods_match_all}
    \end{subfigure}
    \caption{Comparison between ToR classification methods.}\label{fig:comparison}
\end{figure}

\subsection{Evaluation based on Manual Ground Truth}\label{sec:ground_truth_comparison}

In the following subsections, we compare our filtered and unfiltered results with previously published methods that infer ToRs based on AS-level paths and public datasets (see Figure~\ref{fig:comparison}).

For consistency, we employ the same implementation as in~\cite{ToR_AnNet}, utilizing BGP announcements from RouteViews~\cite{RV} (April 2021). The algorithms included in our comparison are:
\begin{enumerate*}
\item GAO – The first ToR inference algorithm, introduced by Gao~\cite{Gao}.
\item SARK – A ToR inference method proposed in~\cite{SARK}.
\item RUAN – An approach introduced by Ruan and Susan Varghese~\cite{ruan} in a technical report.
\item ND-ToR – A near-deterministic AS ToR inference algorithm~\cite{NDToR}, which employs the kmax-Core approach.
\end{enumerate*}
Additionally, we compare our results with two datasets: the CAIDA AS-relationships dataset\cite{caida_asrelationship} and the ProbLink approach\cite{coretoleaf_19}, both from April 2021.

Table~\ref{tab:man_eval_IRR} summarizes the evaluation results of both our filtered and unfiltered results as well as the previous methods based on the {\em manual IRR} dataset. Our Filtered and unfiltered results have the highest overlap with the ground truth (260 and 336 ToRs, respectively), and the filtered results also have the highest accuracy (95.4\%, Ruan achieved 95.3\%). ProbLink had the worst accuracy, 84.5\%, and the smallest overlap, only 174 ToRs.  Problink and SARK accuracy were significantly below all other methods. 

Table~\ref{tab:man_eval_IRR} validates that our approach manages to obtain data with high accuracy of over 95\%.  If the ToR sample we used is representative (and the data supplied in Section~\ref{sec:dataset} suggest that if a bias exists, it is not significant), it is obtained for about 2/3 of the data that can be extracted from the IRR.  If we are ready to lower our accuracy to 92.6\% by using the unfiltered data, our coverage climbs up to about 6/7.  Interestingly, we classified 28 out of the 31 P2P ToRs in the manual dataset; all of them were correctly classified in the filtered dataset.  The CAIDA and Problink datasets classified only 16 and 17, respectively, of these ToRs, each of them had two classification errors, namely, an accuracy of 88\%.  

121 of the links in the unfiltered IRR dataset were not classified by any other method, 106 of them were correctly classified. 87 of the links remained in the filtered list, 82 of them were correctly classified, an improvement in the classification accuracy.

Next, we challenged our algorithm with about 300 manual entries from the BGProtect {\em manual routing} dataset, where the ToR justification was routing (either BGP or traceroute), and IRR was not mentioned. Indeed (see Table \ref{tab:man_eval_route}), our coverage was poor, only 42 entries were left in the filtered dataset, but the accuracy was still 83.3\%.  This dataset demonstrates how the filtering process manages to improve accuracy, the unfiltered dataset has 53 links with an accuracy of 75\%, and the filtering removes 13 links, 6 of them (half) are wrongly classified. This dataset had only 10 P2P links, the filtered dataset classified only two of them, both correctly.  All other datasets had a poor performance for the P2P ToRs. 

Small as it is, 5 of the 53 links in our IRR dataset could not be found in any other dataset, all of them were correctly classified.  4 of them appear in the filtered IRR dataset.

The two manual datasets shows the limitation of using IRR as a single source for ToR classification.  They show that we extract the most of the available information contains in the IRR even when compared to a human analyst: when the ToR is clear in the IRR for an expert, our automatic process classifies it with very high accuracy (over 95\%); but when it doesn't we do not classify in most cases, and when we do it is with reasonably high accuracy (76-83\%).

\begin{figure}[h!]
    \begin{subfigure}[h]{0.5\linewidth}
    \centering
    \includegraphics[width=0.95\linewidth]{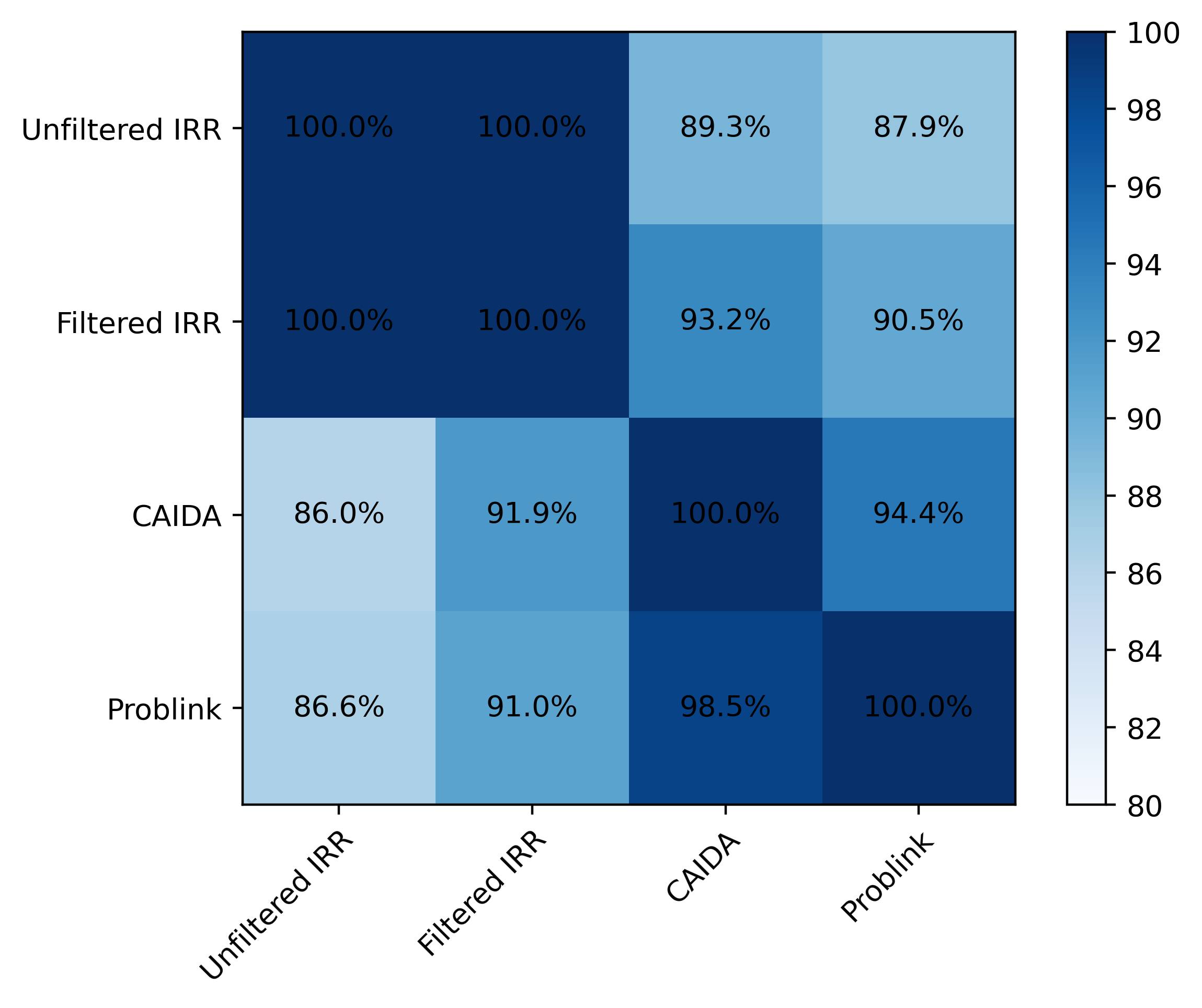}
    \caption{P2P agreement}\label{fig:methods_match_p2p_4}
    \end{subfigure}%
    \begin{subfigure}[h]{0.5\linewidth}
    \centering
    \includegraphics[width=0.95\linewidth]{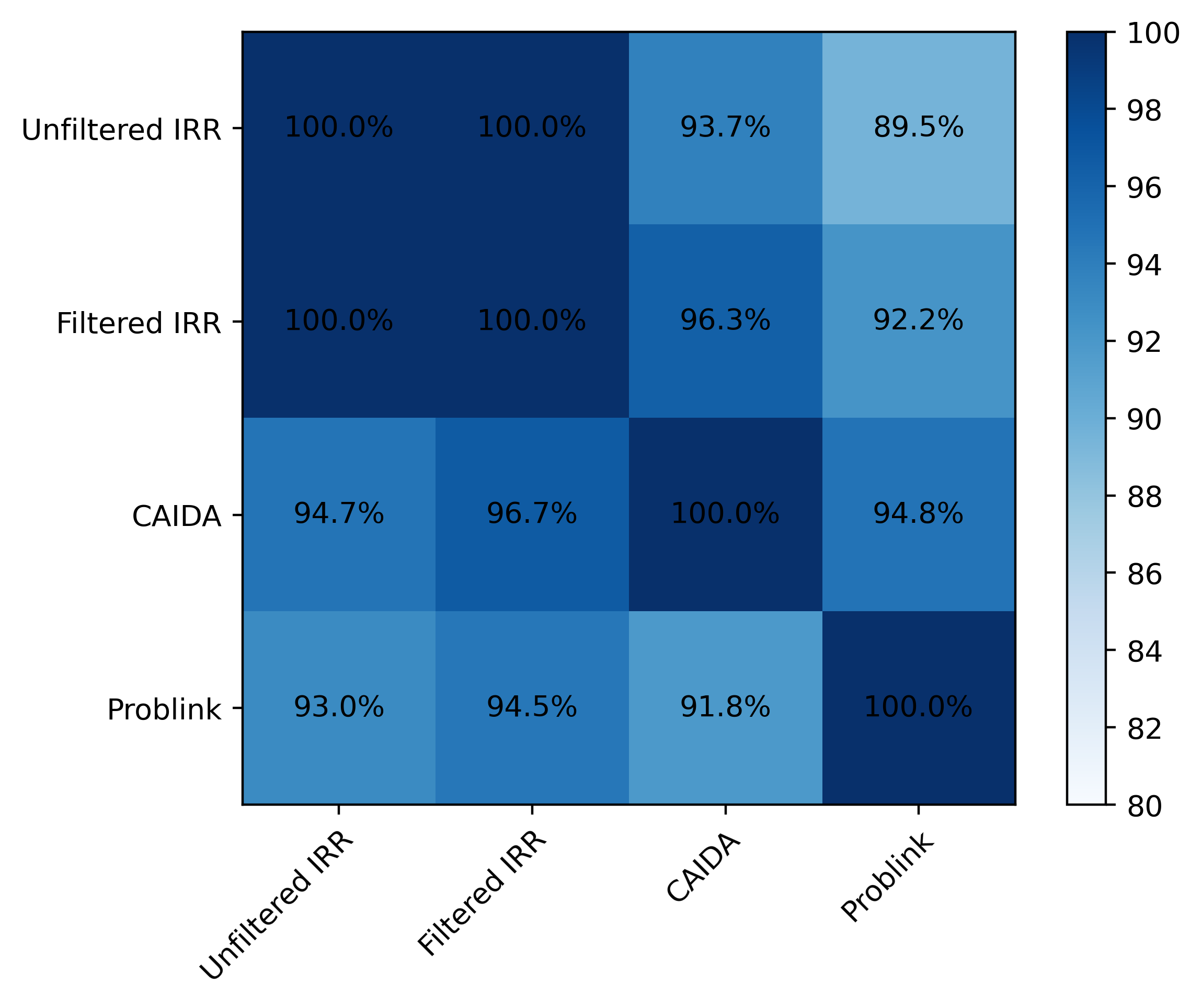}
    \caption{P2C agreement}\label{fig:methods_match_p2c_4}
    \end{subfigure}
    \caption{Cross comparison between our methods, CAIDA and Problink.}\label{fig:cross_comparison}
\end{figure}

\subsection{Cross Comparison} \label{sec:cross}

Figure~\ref{fig:comparison} cross-compares our filtered and unfiltered results with previously published methods. Observing Figure~\ref{fig:comparison} it is clear that our filtering manages to obtain a subset of the links that are classified with high confidence: the agreement (see Eq.~\ref{eq:r}) of our filtered classification with all the other methods is between 89.5\% (GAO and SARK) and 95.5\% (CAIDA). 
%
Interestingly, while there is a high agreement between GAO, RUAN, NDTOR, and SARK (SARK a bit less), and high agreement between CAIDA and ProbLink, the agreement between the two groups is significantly lower (83-87\%). Our IRR results (even the unfiltered results) seem to strike the best matching with both groups.

SARK agreement, as noted above, is a bit worse than the others. 
The reason for this is that SARK is weak in identifying P2P links: it identified correctly only 44\% of our P2P links and only 55.2\% of CAIDA P2P links while the other methods had an agreement above 82.8\% and 89.3\%, respectively.

It is clear from Figure~\ref{fig:comparison} that our results are closely matched with CAIDA and Problink, thus, we further examined the agreement based on our ToRs.
Figure~\ref{fig:cross_comparison} shows the correlation between our ToR classification and CAIDA and ProbLink.  There is a stronger agreement with CAIDA than with Problink for both type of ToRs. There is also a stronger agreement for P2C links, which can be explained by the fact that many P2C links are 'obvious', namely they can be easily deduced, either in the IRR or using other methods.

Some previous works used data inferred from BGP community values as the ground-truth \cite{CAIDA13,coretoleaf_19,TopoScope}. Our comparison to \cite{CAIDA13,coretoleaf_19} can serve as a proxy for this, especially CAIDA series-2.

While our results achieve the best agreement with the other methods, 
the other methods have a larger overlap of more than 150,000 links (except for Problink), while we had an overlap of 59,514 links after filtering. Note again how the filtering process improves the coherence of the results.
The relatively low overlap with the other methods indicates that we managed to enrich the ability to identify ToRs beyond the data mined by the other methods.

\section{Conclusions}

This paper presents ToR classification heuristics, which are solely based on the IRR data.  We produce a large dataset of ToRs, each is associated with a reliability score. 
We showed how to further improve the dataset reliability by filtering and obtained 194,397 P2P and P2C ToRs and 24,743 siblings with very high reliability (148,847 classifications above 90\% reliability). We showed high confidence in our results using a ground-truth dataset of hundreds of ToRs. 
We present how we selected the parameters to achieve a good trade-off between reliability and coverage and provide sufficient details to allow practitioners to select other trade-offs.

We show that about two-thirds of our classified ToRs do not appear in the other datasets or in the RouteViews data used by other methods. In particular, our sibling dataset is an order of magnitude larger than the one published by ProbLink (other methods do not identify siblings).

As a future research direction, we want to classify additional ToRs using machine learning (ML) algorithms based on a test set that contains the ones that we already classified using the IRR heuristics. We can also use ML to identify the small percentage of ToRs that may be wrongly classified.

\section*{Acknowledgment}
\addcontentsline{toc}{section}{Acknowledgment} 

This research was funded in part by the Israeli PMO cyber grant program and by the Blavatnik Interdisciplinary Cyber Research Center at Tel Aviv University.